\documentclass[a4paper,reqno,12pt,final]{amsart}
\usepackage{amsmath,amssymb,eucal,bbold}
\usepackage{array}
\usepackage[T1]{fontenc}
\usepackage[latin1]{inputenc}
\usepackage{graphicx}
\usepackage{subfigure}
\newcommand{\ZZ}{\mathbb{Z}}
\newcommand{\RR}{\mathbb{R}}
\newcommand{\CC}{\mathbb{C}}

\newcommand{\SL}{\mathrm{SL}}
\newcommand{\SU}{\mathrm{SU}}
\newcommand{\SO}{\mathrm{SO}}
\renewcommand{\O}{\mathrm{O}}

\renewcommand{\d}{\partial}
\newcommand{\D}{\mathsf{D}}
\newcommand{\M}{\mathsf{M}}

\newcommand{\ads}{\mathrm{AdS}}
\newcommand{\TT}{\mathbb{T}}
\newcommand{\Ad}{\mathrm{Ad}}
\newcommand{\Aut}{\mathrm{Aut}}
\newcommand{\Inn}{\mathrm{Inn}}
\newcommand{\Out}{\mathrm{Out}}
\newcommand{\eC}{\EuScript{C}}

\newcommand{\eZ}{\EuScript{Z}}
\newcommand{\1}{\mathbb{1}}
\newcommand{\fg}{\mathfrak{g}}
\newcommand{\fn}{\mathfrak{n}}
\newcommand{\half}{\tfrac{1}{2}}

\newcommand{\eN}{\EuScript{N}}
\newcommand{\eE}{\EuScript{E}}
\newcommand{\eS}{\EuScript{S}}
\newcommand{\im}{\mathrm{Im}}
\newcommand{\re}{\mathrm{Re}}
\newcommand{\eA}{\EuScript{A}}
\newcommand{\sP}{\mathsf{P}}
\newcommand{\sJ}{\mathsf{J}}
\newcommand{\sK}{\mathsf{K}}
\begin{document}

\title{More $\D$-branes in the Nappi--Witten background}
\author[Figueroa-O'Farrill]{José Miguel Figueroa-O'Farrill}
\address{\begin{flushright}Department of Physics\\
Queen Mary and Westfield College\\
London E1 4NS, UK\end{flushright}}
\curraddr{
  \begin{center}
    \begin{minipage}[h]{0.6\textwidth}
      \begin{flushleft}
        Department of Mathematics and Statistics\\
        The University of Edinburgh\\
        James Clerk Maxwell Building\\
        The King's Buildings\\
        Mayfield Road\\
        Edinburgh EH9 3JZ, UK
      \end{flushleft}
  \end{minipage}
\end{center}}
\email{jmf@maths.ed.ac.uk}
\author[Stanciu]{Sonia Stanciu}
\address{\begin{flushright}Theoretical Physics Group\\
Blackett Laboratory\\
Imperial College\\
London SW7 2BZ, UK\\
\end{flushright}}
\curraddr{
  \begin{center}
    \begin{minipage}[h]{0.6\textwidth}
      \begin{flushleft}
        Institute for Theoretical Physics\\
        Utrecht University\\
        Princetonplein, 5\\
        3584 CC Utrecht\\
        The Netherlands
      \end{flushleft}
  \end{minipage}
\end{center}}
\email{s.stanciu@phys.uu.nl}
\date{\today}
\thanks{Edinburgh MS-99-005, QMW-PH-99-10, Imperial/TP/98-99/60}
\begin{abstract}
  We re-examine the problem of determining the possible $\D$-branes in
  the Nappi--Witten background.  In addition to the known branes, we
  find that there are also $\D$-instantons, flat euclidean
  $\D$-strings and curved $\D$-membranes admitting parallel spinors,
  all of which can be interpreted as (twisted) conjugacy classes in
  the Nappi--Witten group.
\end{abstract}
\maketitle

\tableofcontents

\section{Introduction and motivation}

$\D$-branes have played a central role in many of the fascinating
developments in string theory in the last few years.  They have proved
particularly versatile in bridging the gap between gauge theory and
gravity.  This is due to the fact that they admit two very different
descriptions: as stable solutions of type II supergravity on the one
hand, and as boundary conditions for open strings on the other.
Despite the fact that both descriptions play equally important roles
in the gravity/gauge theory correspondence, the present state of our
knowledge displays a conspicuous lack of symmetry.  Whereas there has
been much progress in constructing $\D$-brane-type solutions to type
II supergravity, it is only in very few cases that we can actually
ascertain that these solutions describe possible boundary conditions
for open strings.  In particular, very little is known about
$\D$-branes (in the sense of open string boundary conditions) in
nontrivial backgrounds, e.g., in curved space.  This mirrors the fact
that whereas one can write down many supergravity vacua, it is not
clear how to describe string propagation on many of them.  It
therefore might seem unreasonable to expect a stringy description of a
$\D$-brane solution on a background for which string propagation
(without branes) cannot be adequately described.  By the same token,
it is not unreasonable to expect that we should be able to
understand $\D$-branes in those backgrounds for which a conformal
field theory can be written down.  Such backgrounds include flat
spaces, orbifolds, toroidal and Calabi--Yau compactifications, and WZW
and coset models.  In many of these cases, $\D$-branes are fairly well
understood from both geometric and conformal field theoretic points of
view, but in some cases (e.g., the WZW model) a lot less is known than
one might expect.

The purpose of the present paper is to re-examine the problem of
determining the possible conformal invariant Dirichlet boundary
conditions in a WZW model from a geometric perspective, and to
illustrate the results in the case of the WZW model associated to the
Nappi--Witten group \cite{NW}.  This string background is privileged
in that it can be treated exactly as a conformal field theory, and the
geometry is also simple enough to be studied classically.  It also
displays, as we shall see, many of the features of $\D$-branes present
in general WZW models.  The analysis complements the work in
\cite{STDNW} in that we study a different type of gluing conditions
for the chiral currents.

There is growing body of literature on the subject of boundary
conditions in WZW models, and we will not attempt to list all relevant
papers here, except for those which have a direct relation to the
present one.  A fuller comparative discussion of the literature can be
found in \cite{SDnotes}, which explains the geometric
interpretation of the gluing conditions on which this paper is based.

This note is organised as follows.  In Section~\ref{sec:DWZW} we
discuss boundary conditions for WZW models and their relation to the
more familiar gluing conditions on the chiral currents.  The approach
here follows the one in \cite{SDnotes}.  In contrast with that
paper, we restrict ourselves to the case of gluing conditions given by
an automorphism.  The associated boundary conditions are then those
associated to conjugacy classes and twisted generalisations thereof.
In Section~\ref{sec:Dbranes} we illustrate these results to the
Nappi--Witten group and we determine its (twisted) conjugacy classes,
identifying those which can be interpreted as $\D$-branes.  We will
see that among these classes one finds curved membranes, flat
euclidean strings and instantons.  In Section~\ref{sec:conc} we
summarise the results of the paper.

\section{$\D$-branes in WZW models}
\label{sec:DWZW}

The WZW model is the theory of harmonic maps $g:\Sigma \to G$ from a
two-dimensional worldsheet $\Sigma$ to a Lie group $G$ possessing a
bi-invariant metric.  This induces on the Lie algebra $\fg$ of $G$ an
invariant scalar product.  Let the worldsheet have boundary
$\partial\Sigma$.  In the simplest case, $\D$-branes will correspond
to certain submanifolds $B\subset G$ which can be used as boundary
conditions; that is, such that the map $g$ takes the boundary of the
worldsheet to $B$.  In other words, in the presence of a $\D$-brane,
admissible field configurations are those maps $g:\Sigma \to G$ such
that $g:\partial\Sigma \to B$.  Not all submanifolds $B\subset G$ are
allowed: a consistent boundary condition must preserve conformal
invariance.\footnote{Of course, there are more consistency conditions,
  namely the ones coming from demanding the compatibility between the
  open and closed string pictures for processes involving these
  $\D$-branes.  In order to impose these extra conditions, however, it
  is necessary to have the quantum field theory associated to the WZW
  under control, and in particular to know the spectrum.  In this
  paper we will only focus on the consistency conditions coming from
  conformal invariance, or less generally from invariance under the
  current algebra.}

\subsection{Boundary conditions}

The conformal symmetry of the WZW model is a consequence of the affine
symmetry
\begin{equation*}
  g(z,\bar z) \mapsto \Omega(z) g(z,\bar z) \bar\Omega(\bar z)^{-1}~,
\end{equation*}
where $\Omega(z)$ and $\bar\Omega(\bar z)$ are independent maps from
$\Sigma$ to $G$ depending holomorphically and anti-holomorphically,
respectively, on the complex coordinate.  One way to guarantee
conformal invariance of a boundary condition is therefore to preserve
a large enough subgroup of the affine symmetry.

Let us fix one boundary component and choose a local parametrisation
of the worldsheet for which the boundary component in question
coincides with the real axis $z=\bar z$.  Then at the boundary the
affine symmetry becomes
\begin{equation*}
  g\bigr|_{\partial\Sigma} \mapsto \Omega(z)
  g\bigr|_{\partial\Sigma} \bar\Omega(z)^{-1}~.
\end{equation*}

It is easy to envisage submanifolds $B\subset G$ which preserve some
of the affine symmetry; although proving the conformal invariance of
the corresponding boundary conditions might be much more difficult.
We will concentrate here on boundary conditions such that $B$ is a
(twisted) conjugacy class, as this will be born out by our analysis of
the geometry associated to certain gluing conditions.

Let $G$ be a connected Lie group and $g_0\in G$ an element.  Let
$C(g_0)$ denote the conjugacy class of the element $g_0$, defined as
the subset of $G$ with the following elements:
\begin{equation*}
  C(g_0) := \left\{ g g_0 g^{-1} \mid g \in G \right\}~.
\end{equation*}
The conjugacy class of an element $g_0$ is therefore the orbit of that
element under the adjoint action of the group: $\Ad_g : G \to G$,
defined by $\Ad_g (g_0) = g g_0 g^{-1}$.  Each conjugacy class is a
connected submanifold of $G$.  Since every element $g_0$ belongs to
one and only one conjugacy class, $G$ is foliated by its conjugacy
classes.  The leaves of the foliation need not all have the same
topology; in other words, the foliation need not be a fibration.

For compact simple groups a well-known result says that every element
is conjugate to some maximal torus $\TT$.  The Weyl group $W =
N(\TT)/\TT$ further relates elements of the maximal torus.  Therefore
the conjugacy classes are parametrised by the quotient $\TT/W$ of a
maximal torus by the Weyl group.  For example, for $G=\SU(2)$, the
conjugacy classes are parametrised by $S^1/\ZZ_2$, which we can
understand as the interval $\theta\in[0,\pi]$.  The conjugacy classes
corresponding to $\theta=0,\pi$ are points, corresponding to the
elements in the centre of $\SU(2)$: $\pm \1$, whereas the classes
corresponding to $\theta\in(0,\pi)$ are spheres.  If we picture
$\SU(2)$, which is homeomorphic to the 3-sphere, as the one-point
compactification of $\RR^3$ where the sphere at infinity is collapsed
to a point, the foliation of $\SU(2)$ by its conjugacy classes
coincides with the standard foliation of $\RR^3$ by 2-spheres with two
degenerate spheres at the origin and at infinity.  Because of the
degeneration of the limiting spheres the foliation is not a fibration.
This is not surprising, since $S^3$ is a circle bundle over $S^2$ and
not a sphere bundle over $S^1$.

For noncompact groups the problem is more subtle since there is no
longer a notion of maximal torus, even in the semisimple case.  In the
absence of a general theory, one has to treat each case independently.
For example, the conjugacy classes of $\SL(2,\RR)$ can be found in
\cite{Sads3},  in the context of $\D$-branes on $\ads_3$.  In this
paper we will determine the conjugacy classes of the noncompact
nonreductive Lie group $\eN$ introduced by Nappi and Witten in
\cite{NW} and which we term the Nappi--Witten group.

The boundary conditions associated with conjugacy classes preserve an
infinite-dimensional affine symmetry
\begin{equation}
  \label{eq:affsymconjclass}
  g\bigr|_{\partial\Sigma} \mapsto \Omega(z)
  g\bigr|_{\partial\Sigma} \bar\Omega(z)^{-1}~,
\end{equation}
where now $\bar \Omega = \Omega$ at the boundary, but otherwise
arbitrary.

A similar situation obtains in the case of boundary conditions which
say that $B\subset G$ is a shifted conjugacy class: $B = h\, C$ or $B
= C\, h$, where $h\in G$ and $C$ is a conjugacy class.  This also
respects the affine symmetry in \eqref{eq:affsymconjclass}, but acting
on $h^{-1} g$ and $g h^{-1}$ respectively.

A way to generalise this situation is to consider twisted conjugacy
classes.  Every automorphism $r$ of the group gives rise to a twisted
version of conjugacy classes, by considering the orbit of a group
element $h$, say, under the twisted conjugation $h \mapsto g h
r(g)^{-1}$.  If the automorphism is inner, so that there is a group
element $k$ such that $r(g) = k g k^{-1}$, then the orbit under the
twisted conjugation is simply the shifted conjugacy class $k^{-1}
C(gk)$; however when the automorphism is not inner, the orbits are
quite different, as we will see below in the case of the Nappi--Witten 
group.  These boundary conditions are also preserved by the
infinite-dimensional affine symmetry given by
\eqref{eq:affsymconjclass}, but where now $\bar\Omega = r(\Omega)$ at
the boundary.

We will not attempt here to prove directly that all these boundary
conditions preserve conformal invariance.  One may argue that these
boundary conditions seem reasonable because in the more familiar case
of free fields, which we can think as an abelian WZW model,
$\D$-branes are essentially points or planes, which can be interpreted
group-theoretically as (twisted) conjugacy classes.  Instead we will
start from the gluing conditions satisfied by the chiral currents and
under some natural assumptions recover the above boundary conditions.

\subsection{Gluing conditions}

In the same way as the conformal invariance of the WZW model is most
easily proven using the infinite-dimensional affine symmetry obeyed by
the currents, determining whether a boundary condition preserves
conformal invariance is most easily achieved by deriving from it a
condition in terms of the chiral currents.

In the algebraic approach to the WZW model, the fundamental dynamical
variables are the affine currents
\begin{equation*}
  J(z) = - \d g \, g^{-1} \qquad\text{and}\qquad \bar J(\bar z) =
  g^{-1}\, \bar\d g~,
\end{equation*}
where $g=g(z,\bar z)$.  The dynamic content of the WZW model is
encoded in the (anti-)holomorphicity of these currents.  These
currents are Lie algebra valued $(1,0)$- and $(0,1)$-forms on the
worldsheet.  The sign is chosen so that $J(z)$ and $\bar J(\bar z)$
obey isomorphic affine algebras.

In this approach, boundary conditions are expressed as gluing
conditions on the chiral currents.  Let $R:\fg \to \fg$ denote an
invertible linear map in the Lie algebra.  At this moment we are not
assuming any further properties of $R$.  By a \emph{gluing condition}
we mean a relation of the form
\begin{equation}
  \label{eq:gluing}
  \bar J(\bar z)\bigr|_{\partial\Sigma} = R
  J(z)\bigr|_{\partial\Sigma}~.
\end{equation}
The fundamental requirement of a gluing condition is that it should
preserve the conformal algebra at the boundary.  In the WZW model, the 
Virasoro generators are made out of the chiral currents via the
Sugawara construction, which takes the form
\begin{equation*}
  T(z) = \left<J(z),J(z)\right> \qquad\text{and}\qquad
  \bar T(\bar z) = \left<\bar J(\bar z),\bar J(\bar z)\right>~,
\end{equation*}
where $\left<-,-\right>$ is a nondegenerate invariant scalar product
in the Lie algebra $\fg$.\footnote{This need not be the same scalar
  product which appears in the operator product algebra obeyed by the
  chiral currents, or indeed in the WZW lagrangian.  In the case of a
  simple algebra, the difference is simply a familiar renormalisation,
  but in the case of nonsemisimple Lie algebras like that of the
  Nappi--Witten group, the difference between the two scalar products
  is proportional to the Killing form, which is now degenerate.
  Nevertheless, as shown in \cite{FS3}, both of these scalar products
  are nondegenerate and can be used to furnish the corresponding Lie
  group with a bi-invariant metric.}  Conformal invariance then
dictates that $R$ should be an isometry of this scalar product.  We
will see below that under some further assumptions it is natural to
demand that $R$ be a Lie algebra automorphism.

It should be mentioned that the map $R$ need not be constant, it could
depend on the point $g\in G$; in other words, $R$ could be a function
from the group $G$ to the group of isometries of $\fg$.  Indeed, as
shown in \cite{STDNW} via a $\sigma$-model analysis of the Neumann
boundary conditions, in order to obtain a $\D$-brane which fills the
whole group manifold (at least in the case of a noncompact group), one
needs to impose a gluing condition with a nonconstant $R$.  The
implicit restriction to constant $R$ has no conceptual basis; it is
simply a practical one, since it is not known in general how to work
quantum-mechanically with the field $g$, but only with the chiral
currents.

\subsection{The geometry of the gluing conditions}

The geometry of a $\D$-brane is encoded in the boundary conditions
that the fields satisfy.  Since the relation between the gluing
conditions and boundary conditions is in most cases not
straightforward, therein lies the major difficulty in the study of
$\D$-branes on WZW models.  One way to understand where the difficulty
lies is to notice that the gluing conditions relate currents which
take values in the Lie algebra, equivalently in the tangent space to
the Lie group at the identity; whereas in order to derive any
geometric information about the $\D$-brane what one needs are local
conditions on the map $g:\Sigma \to G$ at the boundary of the
worldsheet, which need not be mapped anywhere near the identity in the
group.  This complication is absent for an abelian group, since the
currents are directly related to the coordinates; but for a nonabelian
group the relationship between the gluing conditions and the boundary
conditions is not totally understood.  In what follows we will present
a natural class of solutions related to (twisted) conjugacy classes,
but as shown in \cite{STDNW,Sads3,SDnotes} these are not the only
possible solutions.

Let us start by writing down the gluing conditions \eqref{eq:gluing}
in terms of the map $g:\Sigma \to G$:
\begin{equation*}
  g^{-1} \bar\d g = - R(\d g g^{-1})~,
\end{equation*}
where in this and in many of the following equations we are implicitly 
evaluating both sides of the equation on the boundary.  This relation
is a linear equation in the Lie algebra, which we identify with the
tangent space to the Lie group at the identity.  We can translate it
to a linear equation at the point $g$:
\begin{equation*}
 \bar\d g = - g R(\d g g^{-1})~,
\end{equation*}
where we have abused notation slightly and written the differentials
of left- and right-multiplication as left and right multiplication by
$g$, as for matrix groups.  Introducing $\lambda$ and $\rho$ as the
differential maps associated to left- and right-translations,
respectively, we can rewrite this equation in a more invariant form as 
follows:
\begin{equation}\label{eq:gluing@g}
 \bar\d g = - (\lambda_g \circ R \circ \rho_g^{-1})\, \d g~.
\end{equation}

We seek to interpret this condition as defining a boundary condition
of the form $g:\d\Sigma \to B$ where $B\subset G$ is a submanifold.
We will \emph{assume} that in addition $B$ is nondegenerate, so that
the bi-invariant metric on $G$ restricts non-degenerately to $B$.
We believe this assumption to be physically reasonable.  Let us remark 
that the bi-invariant metric on $G$ is the one induced by the
invariant scalar product in the Lie algebra which is used in the
Sugawara construction.  As remarked earlier this need not agree with,
or be in the same conformal class as, the metric appearing in the WZW
lagrangian and hence in the operator product algebra of the chiral
currents.

The assumption on the submanifold $B\subset G$ says that at any point
$g\in B$ one has the following orthogonal decomposition of the tangent
space to $G$:
\begin{equation*}
  T_g G = T_g B \oplus T_g B^\perp~.
\end{equation*}
We will let ${}^\perp$ denote the orthogonal projection
$T_g G \to T_g B^\perp$ along $T_g B$.  In the open string picture, a
(Dirichlet) boundary condition $g:\d\Sigma \to B$ says that
the normal component of the tangential derivative along the boundary
vanishes, whence
\begin{equation}\label{eq:Dirichlet}
  (\bar\d g)^\perp = - (\d g)^\perp~.
\end{equation}
Because of the bi-invariance of the metric and the fact that $R$ is an 
isometry, the linear map $(\lambda_g \circ R \circ \rho_g^{-1})$
respects the above orthogonal decomposition, whence the normal
component of equation \eqref{eq:gluing@g} together with
\eqref{eq:Dirichlet} implies that
\begin{equation*}
  (\1 - \lambda_g \circ R \circ \rho_g^{-1}) (\d g)^\perp = 0~.
\end{equation*}
Since $\d g$ is arbitrary on the boundary, this is obeyed by
\emph{all} normal vectors to $B$ at $g$, and hence defines what it
means for a vector to be normal to $B$.  This in turn will tell us
what it means for a vector to be tangent to $B$.

The above equation says that a vector $\eta \in T_g G$ is normal to
$B$ if and only if it belongs to the kernel of the linear map
\begin{equation*}
  (\1 - \lambda_g \circ R \circ \rho_g^{-1}) : T_g G \to T_g G~.
\end{equation*}
In other words,
\begin{equation*}
  T_g B^\perp = \ker (\1 - \lambda_g \circ R \circ \rho_g^{-1})~.
\end{equation*}
Hence the tangent space $T_g B$ to $B$, being defined as the orthogonal
complement of $T_g B^\perp$, is then the image of the adjoint of the
above linear map
\begin{equation*}
  T_g B = T_g B^{\perp\perp} = \left[\ker (\1 - \lambda_g \circ R \circ
  \rho_g^{-1})\right]^\perp = \mathrm{im} (\1 - \lambda_g \circ R \circ
  \rho_g^{-1})^\dagger~,
\end{equation*}
where
\begin{multline*}
  (\1 - \lambda_g \circ R \circ \rho_g^{-1})^\dagger =
  \1 - (\lambda_g \circ R \circ \rho_g^{-1})^\dagger\\
  = \1 - (\lambda_g \circ R \circ \rho_g^{-1})^{-1} = \1 - \rho_g
  \circ R^{-1} \circ \lambda_g^{-1}~,
\end{multline*}
where we have used the fact that $(\lambda_g \circ R \circ
\rho_g^{-1})$ is an isometry, whence its adjoint is its inverse.  In
other words, the tangent space $T_g B$ to $B$ at $g$ is made out of
vectors of the form
\begin{equation*}
  \xi - R^{-1}(g^{-1} \xi) g
\end{equation*}
for any $\xi\in T_g G$.  Now, the tangent vectors $\xi$ can be put in
one-to-one correspondence with vectors $X$ in the Lie algebra via the
relation: $\xi = - g R(X)$.  Therefore $T_g B$ consists of vectors of
the form
\begin{equation*}
  X g - g R(X)\qquad\text{for $X\in \fg$.}
\end{equation*}

There is one further condition that we have to impose.  Since
$B\subset G$ is a submanifold, its tangent vectors span an integrable
distribution (in the sense of Frobenius); that is, the Lie bracket of
two tangent vectors should again be a tangent vector.  In other words, 
one must impose that
\begin{equation}\label{eq:frobenius}
  [Xg - gR(X), Yg - gR(Y)] = Zg - gR(Z)\quad\text{for some
    $Z\in\fg$.}
\end{equation}
To compute the above Lie bracket we notice that left-invariant vector
fields generate right-translations, which are anti-homomorphisms,
whereas right-invariant vector fields generate left-translations,
which are homomorphisms, and that left- and right-translations
commute.  Therefore, one computes
\begin{equation*}
  [Xg - gR(X), Yg - gR(Y)] = [XY]g - g[R(X)R(Y)]~.
\end{equation*}
Comparing with the right-hand side of equation \eqref{eq:frobenius} we 
see that
\begin{equation*}
  [XY]g - g[R(X)R(Y)] = Zg - gR(Z) \qquad\text{for some
    $Z\in\fg$.}
\end{equation*}

A natural solution is $Z=[XY]$ and hence $[R(X)R(Y)]=R([XY])$, so
that $R$ is an automorphism; but it is important to realise that,
since there is no unique way of decomposing a vector field into the
sum of a left- and a right-invariant vector fields, this is \emph{not}
the only solution.  Indeed, as evidenced by some of the results in
\cite{STDNW,Sads3,SDnotes}, there are conformally invariant
boundary conditions in WZW models for which $R$ is \emph{not} an
automorphism.

For the purposes of this paper we will however \emph{assume} that $R$
is an automorphism, and discuss the resulting boundary conditions.

Let us offer two remarks:
\begin{itemize}
\item An automorphism $R$ leaves invariant the Killing form and hence
  if it is an isometry with respect to the scalar product in the
  Sugawara construction it will also be an isometry with respect to
  the scalar product appearing in the WZW lagrangian or in the
  operator product algebra of the chiral currents; and
\item Taking $R$ to be automorphism forces $B$ to be a submanifold.
  This means that in order to describe configurations of intersecting
  branes, for example, where $B$ is the union of two submanifolds, one
  is forced to consider gluing conditions where $R$ is not an
  automorphism.
\end{itemize}

\subsection{$\D$-branes and (twisted) conjugacy classes}

We assume then that $R$ is an automorphism, so that the submanifold
$B\subset G$ defining the boundary conditions is such that its tangent
vectors at $g$ are all of the form $Xg - gR(X)$ for some $X\in\fg$.
Such a vector is tangent to the following curve through $g$:
\begin{equation}\label{eq:curve}
  \gamma_X(t) = e^{tX} g e^{-tR(X)}~.
\end{equation}
Let us define the map $r: G\to G$ by 
\begin{equation*}
  r \left( e^{tX} \right) = e^{t R(X)}~,
\end{equation*}
for small enough $t$ and for all $X \in \fg$.  Since we assume that
the group is connected, $r$ extends to a Lie group automorphism.  It
is moreover clear that $r$ is an isometry relative to the bi-invariant
metric on the Lie group provided that $R$ preserves the scalar product
in the Lie algebra.  Therefore the curves described by
\eqref{eq:curve} correspond to curves on the orbit of the point $g$ 
under the twisted adjoint action of the group: $\Ad^r(h) g := h g
r(h)^{-1}$, and hence $B$ can be identified with the orbit of $g$
under such an action:
\begin{equation*}
  B = C_r(g) = \left\{ h g r(h)^{-1} \mid h \in G\right\}~.
\end{equation*}
We call these orbits \emph{twisted conjugacy classes}, since they
reduce to conjugacy classes when $r$ is the identity.\footnote{While
  we were typing the present paper, a paper \cite{FFFS} appeared in
  which one can find the result, obtained using a significantly
  different method, that twisted conjugacy classes are possible
  conformally invariant boundary conditions in WZW models.}

In contrast to the case of conjugacy classes, the twisted adjoint
action does not preserve the group multiplication 
in general.  Nevertheless, the twisted conjugacy class $C_r(g)$ is
again a homogeneous space
\begin{equation*}
  C_r(g) \cong G/H_r(g)~,
\end{equation*}
where $H_r(g) \subset G$ is now the isotropy subgroup of the element
$g$:
\begin{equation*}
  H_r(g) := \left\{ h\in G \mid h g = g r(h)\right\}~.
\end{equation*}

When $R$, and hence $r$, is the identity map, then $B$ agrees with the
conjugacy class of the element $g$.  The result that conjugacy classes
could appear as $\D$-branes appeared for the first time in \cite{AS}.
In \cite{Sads3} (see also \cite{SDnotes}) it was further shown
that in the case where $R$ is an inner automorphism, say $R = \Ad_h$
for some fixed group element $h$, the gluing condition
\eqref{eq:gluing} reads
\begin{equation}
  \label{eq:inner}
  - \d g g^{-1} = h g^{-1} \bar\d g h^{-1}~,
\end{equation}
which is equivalent to
\begin{equation*}
  - \d \tilde g \tilde g^{-1} = \tilde g^{-1} \bar\d\tilde g~,
\end{equation*}
for $\tilde g = g h^{-1}$.  This gluing condition therefore
implies the boundary condition $\tilde g:\d\Sigma \to C$,
where $C$ is a conjugacy class; or equivalently $g: \d\Sigma \to Ch$,
where $Ch$ is the right-translate by $h$ of the conjugacy class $C$.
Notice that $Ch = hC$ so that there is no ambiguity had we chosen to
rewrite the gluing condition in terms of $h^{-1} g$.

More generally two automorphisms which are related by an inner
automorphism give rise to twisted conjugacy classes which are
translated relative to each other:
\begin{equation*}
  C_{r\circ\Ad_h}(g) = C_r(gr(h))r(h)^{-1}\qquad\text{and}\qquad
  C_{\Ad_h\circ r}(g) = C_r(gh)h^{-1}~.
\end{equation*}
This suggests that we organise the possible boundary conditions
according to the group of (metric-preserving) outer automorphisms.
Indeed, let $\Aut _o(G)$ denote the group of metric-preserving
automorphisms of $G$ and let $\Inn_o(G)\subset \Aut_o(G)$ denote the
invariant subgroup corresponding to those automorphisms which are
inner.  Then one can define the factor group
\begin{equation*}
  \Out_o(G) := \Aut_o(G)/\Inn_o(G)~,
\end{equation*}
of metric-preserving outer automorphisms.  Elements of $\Out_o(G)$ are
equivalence classes of metric-preserving automorphisms of $G$: two
such automorphisms being equivalent if they are related by an inner
automorphism.  To each element of $\Out_o(G)$ we can associate an
equivalence classes of $\D$-branes foliating $G$, two such branes
being equivalent if one is simply a translate of another.  For
example, corresponding to the identity in $\Out_o(G)$ we have
conjugacy classes and their translates.  For $G$ a compact simple
group, $\Out_o(G)$ is given by automorphisms of the Dynkin diagram,
which are classified.  For abelian groups no nontrivial automorphism
is inner.  In the case of the Nappi--Witten group, which we treat in
detail in the next section, we have that $\Out_o(G) \cong \ZZ_2$,
whence there will be two distinct families of $\D$-branes, as we will
see below.

It is tempting to interpret the group of metric-preserving outer
automorphisms as a kind of duality group in the WZW model, permuting
different types of $\D$-branes much in the same way as T-duality  or
mirror symmetry in toroidal and Calabi--Yau compactifications,
respectively.

\section{$\D$-branes in the Nappi--Witten group}
\label{sec:Dbranes}

In this section we determine the $\D$-branes in the Nappi--Witten
group corresponding to (twisted) conjugacy classes.  We start with
some general remarks about the Nappi--Witten group we will make use of 
in the sequel.

\subsection{The Nappi--Witten group}

The Nappi--Witten group $\eN$ is the universal central extension of
the two-dimensional euclidean group.  Topologically $\eN \cong S^1
\times \RR^3$, although one sometimes also considers its universal
cover $\Tilde\eN \cong \RR^4$.

It is therefore convenient to parametrise $\eN$ by a triple $(\theta,
w, t)$ where $\theta$ is an angle, $w$ is a complex number and $t$ is
a real number.  A typical group element shall be denoted $g(\theta, w,
t)$.  Our first task is to write down the group multiplication law.
We start with the two-dimensional euclidean group $\eE$, with
elements $h(\theta, w)$.  It acts on the complex plane, parametrised by 
$z$, via affine transformations:
\begin{equation*}
  h(\theta,w) \cdot z = e^{i\theta} \, z + w~.
\end{equation*}
From this action we can read off the group multiplication law:
\begin{equation*}
  h(\theta_1,w_1) \, h(\theta_2,w_2) =
  h(\theta_1+\theta_2, w_1 + e^{i\theta_1}\,w_2)~,
\end{equation*}
which displays the semi-direct product nature of the group.  If we
denote by $R(\theta)$ the rotation by an angle $\theta$ in the complex 
plane: $R(\theta) \cdot z = e^{i\theta}\, z$ and $T(w)$ the
translation by $w$: $T(w) \cdot z = z + w$, then clearly
$h(\theta, w) = T(w) \, R(\theta)$, and the group multiplication
law above says, among other things, that
\begin{align}
  R(\theta_1) \, R(\theta_2) &= R(\theta_1 + \theta_2)
  \label{eq:abelian}\\
  R(\theta) \, T(w) &= T(e^{i\theta}\,w)\, R(\theta)
  \label{eq:semidirect}
\end{align}

Since the group $\eN$ is a central extension of $\eE$, a
representation of $\eN$ is a projective representation of $\eE$.  Such 
a representation is characterised by a group cocycle.  In the present
case, the cocycle is associated with the translations, which as a
result no longer commute:
\begin{equation*}
  T(w_1) \, T(w_2) = T(w_1 + w_2) \, e^{-\half\im(w_1\bar w_2)}~.
\end{equation*}
(The normalisation has been chosen for later convenience.)  Let us
introduce an abstract one-parameter subgroup $Z(t)$ which acts as
$e^t$ in the above projective representation.  Then we can rewrite the
above equation as
\begin{equation}
  \label{eq:cocycle}
  T(w_1) \, T(w_2) = T(w_1 + w_2) \, Z(-\half \im(w_1\bar w_2))~.
\end{equation}

Let $g(\theta, w, t) = T(w)\, R(\theta)\, Z(t)$ denote an element 
of $\eN$.  Then the group multiplication can be read off from
equations \eqref{eq:abelian}, \eqref{eq:semidirect} and
\eqref{eq:cocycle} to give:
\begin{multline}
  \label{eq:grouplaw}
  g(\theta_1, w_1, t_1)\, 
  g(\theta_2, w_2, t_2)\\
  = g(\theta_1+\theta_2, w_1 + e^{i\theta_1}\,w_2 , t_1 + t_2 - \half
  \im(w_1 e^{-i\theta_1} \bar w_2))~.
\end{multline}
It follows from that $g(0,0,0)$ is the identity element and that the
inverse of $g(\theta,w,t)$ is given by
\begin{equation}
  \label{eq:inverse}
  g(\theta,w,t)^{-1} = g(-\theta, - e^{-i\theta}\, w, -t)~.
\end{equation}

\subsection{The Nappi--Witten Lie algebra}
\label{sec:nwalg}

In order to make contact with the traditional approach to the WZW
model, where the fundamental dynamical variables are the chiral
currents, we must discuss the Lie algebra $\fn$ of the Nappi--Witten
Lie group $\eN$.

Let us introduce abstract generators $\sP_1$, $\sP_2$, $\sJ$ and $\sK$
for the Lie algebra $\fn$ of $\eN$.  We postulate the following
relations between these generators and the group elements $T(w)$,
$R(\theta)$ and $Z(t)$:
\begin{equation}
  \label{eq:exp}
  T(w) = \exp(x\sP_1 + y\sP_2)~,\quad R(\theta) = \exp(\theta\sJ)
  \quad\text{and}\quad Z(t) = \exp(t\sK)~,
\end{equation}
where $w = x + i y$.  From the equations \eqref{eq:abelian},
\eqref{eq:semidirect} and \eqref{eq:cocycle} one can easily work out
the Lie brackets obeyed by these generators.  One finds
\begin{equation}
  \label{eq:nwalg}
  [\sJ,\sP_i] = \epsilon_{ij} \sP_j~,\qquad\text{and}\qquad [\sP_i,\sP_j] =
  \epsilon_{ij}\sK~,
\end{equation}
with all other brackets vanishing.  As is well known by now, this
algebra, although solvable, possesses an invariant scalar product with
lorentzian signature:
\begin{equation}
  \label{eq:lametric}
  \langle \sP_i,\sP_j\rangle = \delta_{ij} \qquad\text{and}\qquad
  \langle \sJ,\sK\rangle = 1~;
\end{equation}
and this means that the group $\eN$ possesses a bi-invariant
lorentzian metric.

We will need the metric in order to determine the geometry of the
$\D$-branes, so we compute it now.  By definition, the metric is given
by
\begin{equation*}
  ds^2(g) = \langle g^{-1}dg, g^{-1}dg \rangle~,
\end{equation*}
where $g^{-1}dg$ denotes the left-invariant Maurer--Cartan
$\fn$-valued one-forms, and $\langle-,-\rangle$ is the scalar product
in $\fn$ given by equation \eqref{eq:lametric}.

A simple calculation reveals that
\begin{multline}
  \label{eq:leftMC}
  g^{-1}dg = (\cos\theta dx + \sin\theta dy) \sP_1 + (\cos\theta dy -
  \sin\theta dx) \sP_2 + d\theta \sJ\\
  + (dt - \half xdy + \half ydx) \sK~,
\end{multline}
whence the metric is given by
\begin{equation}
  \label{eq:metric}
  ds^2 = dx^2 + dy^2 + 2 d\theta dt + (ydx - xdy)d\theta~.
\end{equation}
Bi-invariance is easily verified by showing that $ds^2 = \langle
dgg^{-1},dgg^{-1}\rangle$.

Let us remark that one can additionally set $\langle \sJ,\sJ\rangle$
equal to any real number and still have an invariant scalar product,
but there exists a Lie algebra automorphism which sets it back to
zero.  Therefore it represents no real loss in generality to demand
that it be zero from the outset.  In any case, its inclusion would
simply add a term $b d\theta^2$ to the metric \eqref{eq:metric}, where 
$b$ is a real constant.  None of the results we obtain are changed in
any qualitative way by the inclusion of this constant.

\subsection{Automorphisms of the Nappi--Witten group}
\label{sec:autos}

In order to determine the different types of $\D$-branes in the
Nappi--Witten group, we must determine the metric-preserving
automorphisms of the Nappi--Witten group $\eN$.  We will spare the
reader the routine calculation by which one determines these
automorphisms, and sketch the calculation instead.  We work with the
Lie algebra: automorphisms of the Lie group and the Lie algebra are
related by the exponential map, which is a diffeomorphism in this case.
Since the Lie algebra $\fn$ is four-dimensional with a minkowskian
scalar product, the automorphisms of $\fn$ which preserve the scalar
product belong to $\O(3,1)$.  We will refer to them as
\emph{orthogonal automorphisms}.  Determining the group of orthogonal
automorphisms of $\fn$ is thus a linear algebra problem, which can be
easily solved to yield the following.

The orthogonal automorphisms of $\fn$ are parametrised by $\RR^2
\times S^1 \times \ZZ_2$.  Indeed suppose that $(x,y)\in\RR^2$,
$\theta\in S^1$ and $\varepsilon=\pm 1$.  Then the automorphism $R$
corresponding to $(x,y,\theta,\varepsilon)$ is given by
\begin{align*}
  R(\sP_1) &= \varepsilon (\cos\theta \sP_1 - \sin\theta \sP_2) +
  (y\sin\theta - x\cos\theta) \sK\\
  R(\sP_2) &= \sin\theta \sP_1 + \cos\theta \sP_2 - \varepsilon
  (x\sin\theta + y\cos\theta) \sK\\
  R(\sJ) &= x \sP_1 + y \sP_2 + \varepsilon \sJ - \half\varepsilon
  (x^2 + y^2) \sK\\
  R(\sK) &= \varepsilon \sK~.
\end{align*}

It is easy to show that automorphisms for which $\varepsilon=+1$ are
inner, whence the group of orthogonal outer automorphisms is
isomorphic to $\ZZ_2$.  This means that there will be two distinct
families of $\D$-branes: (the translates of) conjugacy classes,
and another family corresponding to (the translates of) twisted
conjugacy classes.  In order to determine these twisted conjugacy
classes, it will prove sufficient to study just one automorphism which
is not inner.  The simplest such automorphism, $R$, is the one
corresponding to $x=y=\theta=0$ and $\varepsilon = -1$:
\begin{equation*}
  R(\sP_1) = -\sP_1 \qquad R(\sP_2) = \sP_2 \qquad R(\sJ) = -\sJ
  \qquad R(\sK) = -\sK~.
\end{equation*}

The induced automorphism $r$ on the Lie group is defined by:
\begin{equation}\label{eq:outer}
r: g(\theta,w,t) \mapsto g(-\theta, - \bar w, -t)~.
\end{equation}
It is easy to check that this is an automorphism which moreover is not
inner, since as we will see below, conjugation leaves $\theta$
invariant.

We now proceed to discuss the two families of $\D$-branes.

\subsection{Conjugacy classes as orbits of the adjoint group}

The first family of $\D$-branes is the one associated to the identity
automorphism, which we have seen give rise to conjugacy classes.

\begin{figure}[h!]
\centering
\mbox{%
\subfigure[$\theta_0=0$]{\includegraphics[width=0.20\textwidth]%
{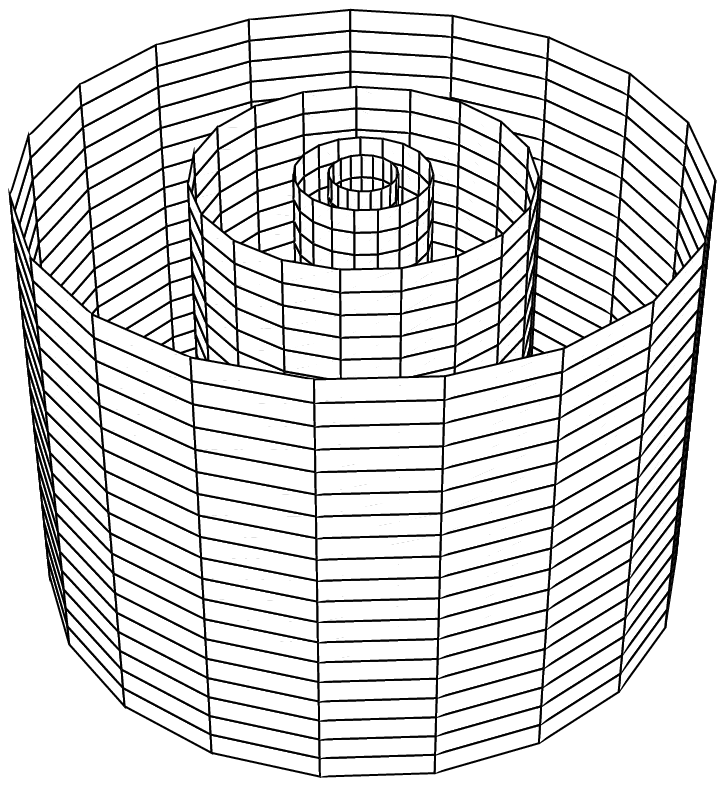}}
\subfigure[$0<\theta_0<\pi$]{\includegraphics[width=0.24\textwidth]%
{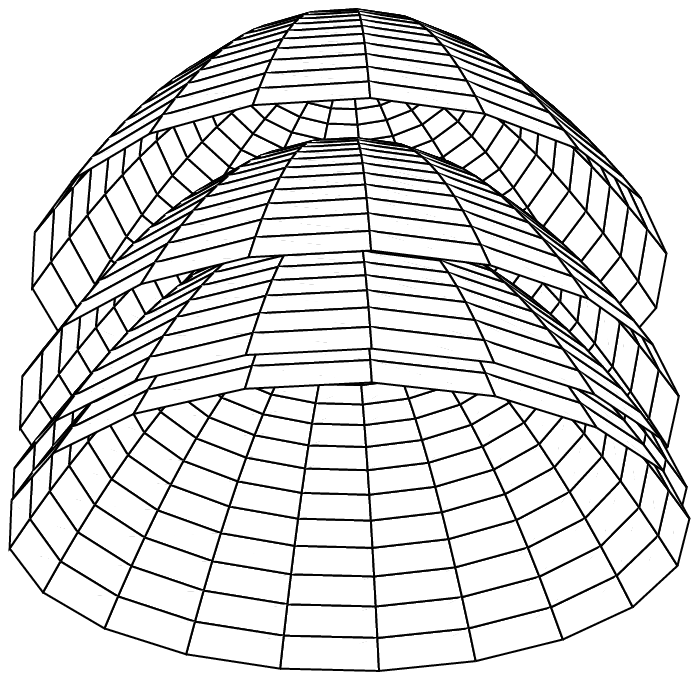}}
\subfigure[$\theta_0=\pi$]{\includegraphics[width=0.20\textwidth]%
{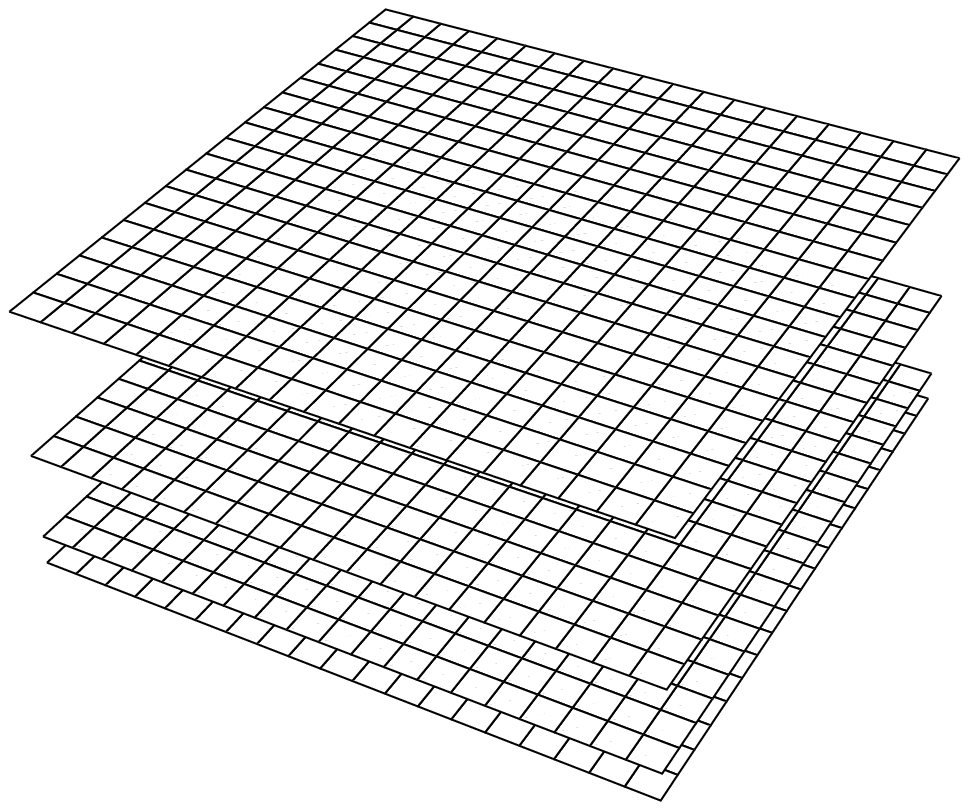}}
\subfigure[$2\pi>\theta_0>\pi$]{\includegraphics[width=0.25\textwidth]%
{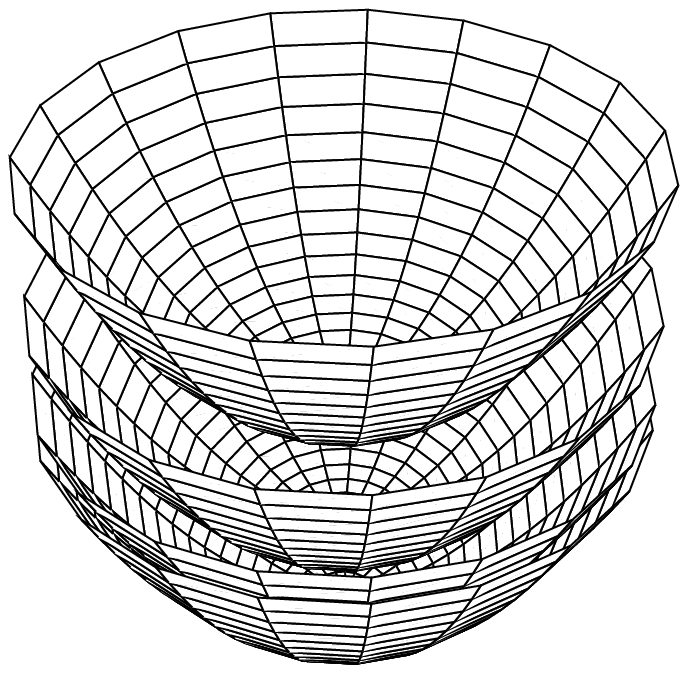}}
}
\caption{Foliation of $\theta_0=\text{constant}$ three-planes by
conjugacy classes.}
\label{fig:Dbranes}
\end{figure}

Using the explicit expression \eqref{eq:inverse} for the inverse of a
group element and the group multiplication law \eqref{eq:grouplaw}, we 
can compute the adjoint action of a generic element $g(\theta, w, t)$
on a fixed element $g(\theta_0, w_0, t_0)$, defined by
\begin{equation*}
  \Ad_{g(\theta, w, t)} \, g(\theta_0, w_0, t_0) := 
  g(\theta, w, t) \, g(\theta_0, w_0, t_0) \, g(\theta, w, t)^{-1}~.
\end{equation*}
We find
\begin{multline}
  \label{eq:adjoint}
  \Ad_{g(\theta, w, t)} \, g(\theta_0, w_0, t_0) =\\
  g\left(\theta_0, (1-e^{i\theta_0})\, w + e^{i\theta}\, w_0, t_0
    -\half |w|^2 \sin\theta_0 -\half \im(1+e^{i\theta_0}) w
    e^{-i\theta} \bar w_0\right).
\end{multline}
The first thing we notice is that $\theta_0$ is an invariant of the
conjugacy class, and we can try to distinguish between the different
conjugacy classes by the value of $\theta_0$ in the first place.  The
three-plane defined by fixing the value of $\theta$ is foliated by the 
conjugacy classes.  The results are summarised in
Table~\ref{tab:Dbranes} and illustrated in
Figure~\ref{fig:Dbranes}.

\begin{table}[h!]
\centering
\setlength{\extrarowheight}{3pt}
\begin{tabular}{|>{$}c<{$}|>{$}c<{$}|c|>{$}c<{$}|c|}\hline 
\text{Class} & \text{Element} & Type & \text{Centraliser} &
Causal Type\\
\hline\hline
\eC'_{0,t_0} & g(0,0,t_0) & point & \eN & \\
\eC_{0,|w_0|} & g(0, w_0, 0) & cylinder & g(0,s w_0, t) & degenerate\\
\eC_{\pi,t_0} & g(\pi,0,t_0) & plane & g(\theta,0,t) & spacelike\\
\eC_{\theta_0\neq0,\pi,~k_0} & g(\theta_0,0,k_0) & paraboloid &
g(\theta,0,t) & spacelike\\
\hline
\end{tabular}
\vspace{8pt}
\caption{Conjugacy classes of the Nappi--Witten group, along with the
  typical element, the topological type of the conjugacy class, the 
  centraliser of the typical element and the causal type of the
  class.}
\label{tab:Dbranes}
\end{table}

Consider first of all the case of $\theta_0=0$.  Then we see that
under the adjoint action \eqref{eq:adjoint},
\begin{equation*}
  g(0,w_0,t_0) \mapsto g\left(0, e^{i\theta}\, w_0, t_0 - \im w
  e^{-i\theta} \bar w_0 \right)~.
\end{equation*}
We can distinguish two subcases:
\begin{itemize}
\item ($w_0=0$) In this case, $g(0,0,t_0)$ belongs to the centre of
  $\eN$ and hence is the only element in its conjugacy class
  $\eC'_{0,t_0}$.  The centraliser of every such class is the group
  $\eN$ itself.
\item ($w_0\neq 0$) In this case, the conjugacy class is a cylinder
  (i.e., diffeomorphic to $\RR \times S^1$) comprising those group
  elements of the form $g(0,w_0 e^{i\theta}, t)$, which shows that
  the conjugacy class is labelled by $|w_0|$.  We denote this class by
  $\eC_{0,|w_0|}$. The centraliser of a typical element $g(0,w_0,0)$
  in this class is the two-dimensional abelian subgroup $\eA_{w_0}
  \subset \eN$ whose elements are of the form $g(0,s w_0, t)$, with
  $s$ and $t$ real numbers.
\end{itemize}
Figure~\ref{fig:Dbranes}(a) shows how these conjugacy classes
foliate the three-plane in $\eN$ corresponding to $\theta_0=0$.

Consider now $\theta_0=\pi$.  In this case, under the adjoint action
\eqref{eq:adjoint},
\begin{equation*}
  g(\pi,w_0,t_0) \mapsto g(\pi, 2w + e^{i\theta} w_0, t_0)~.
\end{equation*}
This shows that the conjugacy class is a 2-plane labelled by $t_0$,
denoted $\eC_{\pi,t_0}$ and pictured in
Figure~\ref{fig:Dbranes}(c).  The centraliser of an element
$g(\pi,w_0,t_0)$ in $\eC_{\pi,t_0}$ is the two-parameter subgroup
$\eS_{w_0}\subset\eN$ consisting of elements of the form
\begin{equation*}
g(\theta, \half (1-e^{i\theta}) w_0, t)~.
\end{equation*}
Topologically, $\eS_{w_0} \cong \RR \times S^1$; although as Lie
groups it is not a product.  Indeed, if we let $k(\theta,t) :=
g(\theta, \half (1-e^{i\theta}) w_0, t)$, then we have
\begin{multline*}
  k(\theta_1,t_1) \, k(\theta_2,t_2)\\
  = k\left(\theta_1 + \theta_2, t_1 + t_2 - \tfrac18 |w_0|^2 \im
    (1-e^{i\theta_1}) (1-e^{i\theta_2})\right)~.
\end{multline*}

Finally we consider the case of general $\theta_0\neq 0,\pi$.  The
centraliser of an element $g(\theta_0, w_0, t_0)$ with $\theta_0\neq
0,\pi$ is given by those elements of the form
\begin{equation*}
  g(\theta,  (1-e^{i\theta}) w_0 (1-e^{i\theta_0})^{-1}, t)~,
\end{equation*}
which form a two-dimensional subgroup $\eS_{z_0} \subset \eN$ with
$z_0:= w_0 (1-e^{i\theta_0})^{-1}$.  The conjugacy class, being
diffeomorphic to $\eN/\eS_{z_0}$, is therefore two-dimensional.  The
conjugacy class is defined (having fixed $\theta_0$) by the value of
the real-valued class function
\begin{equation*}
  F(\theta_0,w_0,t_0) := t_0\, \sin\tfrac{\theta_0}{2} + \tfrac14
  |w_0|^2\, \cos\tfrac{\theta_0}{2}~.
\end{equation*}
Fixing a real number $k_0$ and $\theta_0\neq 0,\pi$, elements in the
conjugacy class $\eC_{\theta_0,k_0}$ are of the form
\begin{equation*}
  g(\theta_0, w, k_0 - \tfrac14 |w|^2 \cot\tfrac{\theta_0}{2})~,
\end{equation*}
for $w\in\CC$, which make up a paraboloid.  We can distinguish two
cases: $\theta_0\in(0,\pi)$ and $\theta_0\in(\pi,2\pi)$.  They are
illustrated in Figures~\ref{fig:Dbranes}(b) and
\ref{fig:Dbranes}(d), respectively.

Parenthetically, the function $F$ also distinguishes conjugacy classes
when $\theta_0=\pi$ and even for $\theta_0=0$ when $|w_0|\neq 0$;
whereas for $w_0=0$, it is the value of $t_0$ which now distinguishes
the (pointlike) conjugacy classes.

\subsection{Geometry of the conjugacy classes}

Due to their interpretation as $\D$-branes, an important
characteristic of a conjugacy class is its geometry.  In particular we
have seen above that in order to conclude that a conjugacy class may
serve as Dirichlet boundary conditions for a WZW model, it was
necessary to assume that it was non-degenerate relative to the
bi-invariant metric.  In this section we will elucidate the geometry
of the conjugacy classes found above.  We will see that the planar and
paraboloidal conjugacy classes are flat and euclidean, whereas the
cylindrical conjugacy class is degenerate.  Therefore all but the
cylindrical conjugacy classes can be understood as $\D$-branes, at
least at our current level of understanding.  Since the metric on
$\eN$ is bi-invariant, the calculation is simplified in that it is
enough to determine the geometry at a point: all other points being
related by the adjoint action of the group, which is an isometry.  At
the same time, left- or right-translating a conjugacy class does not
alter its geometry.

As alluded to above, the bi-invariant metric on $\eN$ is not unique
(even up to scale): one can always add a term $b d\theta^2$ for
$b\in\RR$.  Since, as we shall see, the conjugacy classes have
constant $\theta$, the induced metric is impervious to the inclusion
of this term.

\subsubsection{The conjugacy classes $\eC_{0,|w_0|}$}

From the above considerations it is enough to work at a point, which
we take to be the typical element $g_0 := g(0,w_0,0)$.  The conjugacy
class is the set $\eC = \{g(0,e^{i\theta}w_0,t)\mid \theta,t\in
\RR\}$, and the centraliser is the subgroup $\eZ = \{ g(0,sw_0,t) \mid
s,t\in\RR\}$.  At the point $g_0$, the tangent space $T_{g_0}\eC$ to
the conjugacy class is spanned by the vectors $v_1 = x_0\d_y -
y_0\d_x$ and $v_2 = \d_t$, where $w_0 = x_0 + i y_0$.  Computing their
scalar product, we find that at the point $g(0,w_0,0)$,
\begin{equation*}
  \langle v_1,v_1 \rangle = |w_0|^2~, \langle v_1,v_2\rangle = \langle 
  v_2,v_2\rangle = 0~.
\end{equation*}
Therefore the metric is degenerate, and this means that this conjugacy
class cannot be interpreted (at least straightforwardly) as a
$\D$-brane.  Notice that the tangent space $T_{g_0}\eZ$ to the
centraliser $\eZ$ at the point $g_0$, is spanned by the vectors $u_1 =
x_0\d_x + y_0\d_y$ and $u_2 = \d_t$, which obey
\begin{equation*}
  \langle u_1,u_1 \rangle = |w_0|^2~, \langle u_1,u_2\rangle = \langle 
  u_2,u_2\rangle = 0~.
\end{equation*}
As usual we find that at $g_0$, $T_{g_0}\eC =
\left(T_{g_0}\eZ\right)^\perp$, but that in this case $T_{g_0}\eC \cap
T_{g_0}\eZ \neq \{0\}$.

\subsubsection{The conjugacy classes $\eC_{\pi,t_0}$}

Let us now fix the typical element $g_0 := g(\pi,0,t_0)$.  The
conjugacy class of this element is the set $\eC = \{g(\pi,w,t_0) \mid
w \in \CC\}$ and the normaliser of the typical element is the subgroup
$\eZ = \{g(\pi+\theta, 0, t_0 + t) \mid \theta,t\in\RR\}$.  At the
point $g_0$, the tangent space to the conjugacy class is spanned by
the vectors $v_1 = \d_x$ and $v_2 = \d_y$, with metric
\begin{equation*}
  \langle v_1,v_1 \rangle = \langle v_2,v_2 \rangle = 1~, \langle
  v_1,v_2\rangle = 0~.
\end{equation*}
This metric is clearly non-degenerate and euclidean.  It is also
evidently flat.  Therefore the resulting $\D$-brane is a flat
euclidean $\D$-string.  The tangent space to the centraliser subgroup
at the point $g_0$ is spanned by $u_1 = \d_\theta$ and $u_2 = \d_t$,
whose metric is
\begin{equation*}
  \langle u_1,u_2 \rangle = 1~, \langle u_1,u_1\rangle = \langle
  u_2,u_2\rangle = 0~,
\end{equation*}
which is non-degenerate and minkowskian.  Again we have that
$T_{g_0}\eC =\left(T_{g_0}\eZ\right)^\perp$, but now $T_{g_0}\eC \cap
T_{g_0}\eZ = \{0\}$~, so that $T_{g_0}\eN = T_{g_0}\eC \oplus
T_{g_0}\eZ$.

\subsubsection{The conjugacy classes $\eC_{\theta_0,k_0}$}

We take $g_0 := g(\theta_0,0,k_0)$ as typical element.  Its conjugacy
class is the set $\eC = \{ g(\theta_0,w,k_0 - \tfrac14 |w|^2 \cot
\tfrac{\theta_0}{2})\mid w\in\CC\}$, and its centraliser subgroup is
$\eZ = \{g(\theta,0,t)\mid \theta,t\in\RR\}$.  The tangent space to
the conjugacy class at $g_0$ is spanned by the vectors $v_1 =\d_x$ and
$v_2 = \d_y$, with metric
\begin{equation*}
  \langle v_1,v_1 \rangle = \langle v_2,v_2 \rangle = 1~, \langle
  v_1,v_2\rangle = 0~.
\end{equation*}
This metric is clearly non-degenerate, euclidean and flat.  Therefore
the resulting $\D$-brane is again a flat euclidean $\D$-string.  The
tangent space to the centraliser subgroup at the point $g_0$ is
spanned by $u_1 = \d_\theta$ and $u_2 = \d_t$, whose metric is
\begin{equation*}
  \langle u_1,u_2 \rangle = 1~, \langle u_1,u_1\rangle = \langle
  u_2,u_2\rangle = 0~,
\end{equation*}
which is non-degenerate and minkowskian.  Again we have that
$T_{g_0}\eC =\left(T_{g_0}\eZ\right)^\perp$, and $T_{g_0}\eC \cap
T_{g_0}\eZ = \{0\}$~, so that $T_{g_0}\eN = T_{g_0}\eC \oplus
T_{g_0}\eZ$.

\subsection{Twisted conjugacy classes of the Nappi--Witten group}

In this section we determine the twisted conjugacy classes in the
Nappi--Witten group.  As discussed above, they will all be translates
of the twisted conjugacy classes corresponding to the automorphism $r$
defined in equation \eqref{eq:outer}, which is a representative for
the unique nontrivial (metric-preserving) outer automorphism in the
Nappi--Witten group.

Thus let $C_r(\theta_0,w_0,t_0)$ denote the twisted adjoint orbit of the
element $g(\theta_0,w_0,t_0)$:
\begin{multline*}
  C_r(\theta_0,w_0,t_0)\\
  := \left\{ g(\theta,w,t) g(\theta_0,w_0,t_0) g(-\theta, - \bar w,
    -t)^{-1} \mid (\theta,w,t) \in S^1\times\CC\times\RR \right\}~.
\end{multline*}

Using the group multiplication law \eqref{eq:grouplaw}, one can
compute the twisted action.  First of all notice that
\begin{equation*}
  g(-\theta, - \bar w, -t)^{-1} = g(\theta, e^{i\theta} \bar w, t)~, 
\end{equation*}
and therefore
\begin{multline}\label{eq:twistedaction}
  g(\theta,w,t) g(\theta_0,w_0,t_0) g(\theta, e^{i\theta} \bar w, t) =
  g(\theta_0 + 2\theta, e^{i\theta} w_0 + w +
  e^{i(2\theta+\theta_0)} \bar w, \\
  \hfill t_0 + 2t - \half\im(we^{-i\theta}\bar w_0) - \half
  \im((w+e^{i\theta}w_0) e^{-i(\theta_0 + 2\theta)} w))~.
\end{multline}

In order to determine the dimension of this orbit, we work out the
isotropy of the point $(\theta_0,w_0,t_0)$; that is, the subgroup of
$\eN$ whose twisted adjoint action leaves $g(\theta_0,w_0,t_0)$
invariant.  Equating the right-hand side of equation
\eqref{eq:twistedaction} with $g(\theta_0,w_0,t_0)$ we find
that $\theta=0$, $w = i \varrho e^{i\theta_0/2}$, and $t = \half
\varrho \re(w_0 e^{-i\theta_0/2})$,  for some real parameter
$\varrho$.  In other words, the isotropy subgroup is one-dimensional
and hence the twisted conjugacy classes are three-dimensional.

Since they have codimension one, it is possible to exhibit these
twisted conjugacy classes as level sets of a function which is
invariant under the twisted adjoint action.  The infinitesimal
generators of the twisted adjoint action are easy to work out and
demanding that a function be invariant under their action gives a
number of partial differential relations.  The differential ring of
functions satisfying these relations is generated by the function
\begin{equation*}
  G(\theta_0,w_0,t_0) = \im( w_0 e^{-i\theta_0/2})~.
\end{equation*}
Hence each twisted conjugacy class is determined by a real number
$k$, as the set of $g(\theta,w,t)$ for which
\begin{equation}\label{eq:levelset}
  \im (w e^{-i\theta/2}) = k~.
\end{equation}

It may be worth remarking that the twisted conjugacy classes in this
family are odd-dimensional.  This is in sharp contrast with the case
of conjugacy classes, which (at least in groups admitting bi-invariant
metrics) always have even dimension.  This follows because conjugacy
classes are the image under the exponential map of the adjoint orbits,
which are diffeomorphic to the co-adjoint orbits which are symplectic
manifolds relative to the natural Kirillov--Kostant--Souriau
symplectic structure, and hence are even-dimensional.

\subsection{Geometry of the twisted conjugacy classes}

The induced metric on the twisted conjugacy classes can be worked out
as follows.  Let us introduce polar coordinates $(r,\phi)$ related to
$w$ in the usual way: $w = r e^{i\phi}$.  Let us assume that $k\neq
0$.  We can then eliminate the radial coordinate $r$ using equation
\eqref{eq:levelset}:
\begin{equation*}
  r = k \csc (\phi-\half\theta)~.
\end{equation*}
It is convenient to eliminate $\phi$ in terms of the real variable $z
= \cot(\phi - \half \theta)$.  In terms of the coordinates $(z,
\theta, t)$ parametrising the twisted conjugacy class, the metric
\eqref{eq:metric} becomes
\begin{equation}
  \label{eq:metricTCC}
  ds^2 = k^2 dz^2 + 2 d\theta\left( dt - \tfrac18 k^2 (1+z^2)
    d\theta\right)~,
\end{equation}
which is clearly non-degenerate and of lorentzian signature.
Therefore twisted conjugacy classes define $\D$-membranes.  Unlike the
euclidean $\D$-strings associated to the conjugacy classes, these
membranes are not flat and, since in three dimensions, Ricci-flatness
implies flatness, they are not Ricci-flat either.  To see this let us
simply notice that the riemannian connection is given by
\begin{equation*}
  \nabla_\theta \d_\theta = \tfrac14 z \d_z
  \qquad\text{and}\qquad
  \nabla_\theta \d_z = \nabla_z \d_\theta = -\tfrac14 k^2 z \d_t~,
\end{equation*}
whence the riemannian curvature tensor has components
\begin{equation*}
  R_{z\theta\theta}{}^z = -\tfrac14 
  \qquad\text{and}\qquad
  R_{z\theta z}{}^t = \tfrac14 k^2~.
\end{equation*}
Notice that the Ricci tensor has non-vanishing components
$R_{\theta\theta} = \tfrac14$, but that the scalar curvature does
vanish.

Let us remark that the membrane admits parallel spinors.  In fact, it
is an example of an indecomposable Ricci-null lorentzian manifold (see
for example \cite{JMWaves}) with holonomy group isomorphic to $\RR
\subset \SO(2,1)$.  In fact the membrane metric conforms to the metric 
given by equation (17) in \cite{JMWaves}, which is the most general
three-dimensional lorentzian metric admitting parallel spinors.

The same conclusion can be reached for $k=0$; although the details are 
different.  In this case, we do not eliminate $r$, but rather notice
that in this case $\phi = 2\theta$, whence in terms of the real
coordinates $(r,\theta,t)$, the induced metric on the membrane is
given by
\begin{equation}
  \label{eq:metricTCCk=0}
  ds^2 = dr^2 + 2d\theta\left( dt - \tfrac38 r^2 d\theta\right)~,
\end{equation}
which again describes an indecomposable three-dimensional lorentzian
manifold admitting parallel spinors.

Let us remark that the induced metric on the twisted conjugacy classes 
is affected by including the term $b d\theta^2$ coming from the
ambiguity in the invariant scalar product $\fn$; but this does not
change qualitatively the geometry of the brane.  One still obtains
indecomposable three-dimensional lorentzian manifolds admitting
parallel spinors.

In summary, we see that the Nappi--Witten group admits curved
membranes admitting parallel spinors, flat euclidean strings
and instantons among its $\D$-branes.

\section{Conclusions}
\label{sec:conc}

In this paper we have re-examined the possible boundary conditions in
a WZW model which are compatible with conformal invariance.  We have
shown (following \cite{SDnotes}) how to relate gluing conditions
on the chiral currents to geometric boundary conditions of the group
valued fields in the classical description of the WZW model.  We have
seen that a natural solution to the consistency conditions are given
by (twisted) conjugacy classes: the orbits of group elements under the 
adjoint action twisted by an automorphism.  When the automorphism is
trivial, the resulting orbits are conjugacy classes, and when it is
inner the orbits are shifted conjugacy classes.  More generally, two
automorphisms whose ``difference''  is an inner automorphism give rise 
to orbits which are shifted relative to each other.  This suggests
that one should use the group of (metric-preserving) outer
automorphisms as a sort of classifying group for boundary conditions.

It is important to keep in mind that not all boundary conditions are
given by (twisted) conjugacy classes, as shown in
\cite{STDNW,Sads3,SDnotes}.

We have then illustrated these results in the case of the
Nappi--Witten group.  In that case the group of (metric-preserving)
outer automorphisms has order 2, and hence there are two distinct
families of possible boundary conditions.  One family gives rise to
(shifted) conjugacy classes.  Studying their geometry, we have seen
that these classes consist of flat euclidean $\D$-strings and
$\D$-instantons.  The other family of boundary conditions gives rise
to $\D$-membranes which are not flat: they are indecomposable
lorentzian manifolds admitting parallel spinors.

%
%

\providecommand{\bysame}{\leavevmode\hbox to3em{\hrulefill}\thinspace}

\end{document}